\documentclass{JHEP3} 

\usepackage{epsfig}
\usepackage{amsmath, amssymb}
\usepackage{concmath, palatino}
\usepackage{mathrsfs}
\usepackage{color}

\newcommand\fverb{\setbox\pippobox=\hbox\bgroup\verb}
\newcommand\fverbdo{\egroup\medskip\noindent%
			\fbox{\unhbox\pippobox}\ }
\newcommand\fverbit{\egroup\item[\fbox{\unhbox\pippobox}]}
\newbox\pippobox

\newcommand{\be}{\begin{equation}}
\newcommand{\ee}{\end{equation}}
\newcommand{\ba}{\begin{eqnarray}}
\newcommand{\ea}{\end{eqnarray}}

\newcommand{\ads}{AdS_5\times S^5}

\newcommand{\ddb}{{\overline{\mathscr D}}}
\newcommand{\sym}{$\mathcal{N}=4$ SYM }

\newcommand{\Z}{\mathbb{Z}}
\newcommand{\N}{\mathcal{N}}

\allowdisplaybreaks

    \newcommand{\eq}[1]{(\ref{#1})}

    \newcommand{\beq}{\begin{equation}}
    \newcommand{\eeq}{\end{equation}}
    \newcommand\beqa{\begin{eqnarray}}
    \newcommand\eeqa{\end{eqnarray}}

\title{Y-system for $\Z_{S}$ Orbifolds of $\N$ = 4 SYM}

\author{Matteo Beccaria\\
  Dipartimento di Fisica, Universita' del Salento, 
  Via Arnesano, 73100 Lecce \&\\
  INFN, Sezione di Lecce\\
  E-mail: \email{matteo.beccaria$\bullet$le.infn.it}
}

\author{Guido Macorini\\
  Dipartimento di Fisica, Universita' del Salento, 
  Via Arnesano, 73100 Lecce \&\\
  INFN, Sezione di Lecce\\
  E-mail: \email{guido.macorini$\bullet$le.infn.it}
}

\abstract{
We propose a twisted Y-system for the calculation of leading wrapping corrections to physical states of 
general $\Z_{S}$ orbifold projections of $\N=4$ super Yang-Mills theory. Agreement with available 
thermodynamical Bethe Ansatz results is achieved in the non supersymmetric case. 
Various examples of new computations, including other supersymmetric orbifolds are illustrated.
}

\begin{document} 

\section{Introduction}
\label{sec:intro}

The study of orbifolds of AdS/CFT duality is a very interesting topic where supersymmetry can be broken  
in a mild way, namely by boundary conditions~\cite{Kachru:1998ys,Lawrence:1998ja}. Planar integrability is known to be preserved \cite{Beisert:2005he,Astolfi:2006is}
and the study of ubiquitous wrapping corrections \cite{Ambjorn:2005wa} is 
clearly interesting.

\medskip
The standard setup is that of  a stack of $N$ D3-branes located on the fixed point of the orbifold $\mathbb{C}^{3}/\Gamma$. By gravitational backreaction, the near-horizon geometry is $AdS_{5}\times (S^{5}/\Gamma)$,
where the discrete group $\Gamma$ is a finite subgroup of $SO(6)$, the isometry group of $S^{5}$. AdS/CFT predicts duality between type II string and a particular {\em orbifold projection} of four dimensional $\N=4$ SYM, where the action of $\Gamma$ is extended on $SU(4)$, the spin cover of $SO(6)$. 
The center of $SU(4)$ is $\Z_{4}$ and there is a $\Z_{2}\equiv \widetilde\Z_{2}\subset \Z_{4}$ such that 
$SU(4)/\widetilde\Z_{2} = SO(6)$. The group $\Gamma$, as a subgroup of $SU(4)$ 
does not contain the non trivial element of $\widetilde\Z_{2}$. 
Thus $\Gamma/\widetilde\Z_{2} \equiv \Gamma$~\cite{Kachru:1998ys,Lawrence:1998ja}. 

\medskip
Remarkably, since $AdS_{5}$ is untouched, one expects large $N$ (or better said, planar) conformal invariance. Actually, the construction is more general and can be applied in the reverse, starting with a generic 
$\Gamma\subset SU(4)$. In this case, one can obtain type 0 string theory if $\Gamma$
does not project trivially on $SO(6)$, {\em i.e.} $\Gamma/\widetilde\Z_{2}\neq \Gamma$ \cite{Klebanov:1998yya,Klebanov:1999ch,Tseytlin:1999ii,Nekrasov:1999mn}.

\medskip
In this paper, we propose a method to compute the leading wrapping finite size corrections  to the abelian
$\Z_{S}$ orbifolds of the $\ads$ superstring \cite{Beisert:2005he}. The most powerful and general approach 
to the evaluation of these corrections is the mirror thermodynamic Bethe Ansatz (TBA) developed for the unorbifoldized theory
in~\cite{TBA1} and deeply tested in \cite{TBA2}, mainly in the theoretical laboratory  of $\mathfrak{sl}(2)$ states. 
The associated Y-system has been conjectured in \cite{GKV} based on symmetry arguments and educated guesses
about the analyticity and asymptotic properties of the Y-functions.

\medskip
The same methods can be extended to study finite-size corrections in theories which are closely related to 
\sym by AdS/CFT duality. These are obtained as ($\beta$-) deformations or quotients of the $\ads$ 
string background.
The crucial ingredient to recover the twisted Bethe equations of \cite{Beisert:2005if}
are the associated deformed transfer matrices, which can be obtained by twisting the undeformed transfer matrix \cite{Arutyunov:2010gu}, or by twisting the S-matrix \cite{Ahn:2010ws}. 
Successful applications to $\beta$-deformed theories have been presented
in \cite{Gromov:2010dy,Beccaria:2010kd,deLeeuw:2010ed} at the level of the transfer matrix, and 
in \cite{Ahn:2010ws,Ahn:2010yv} in the $S$-matrix formalism.

\medskip
Analogous applications to the case of orbifolds has been presented in \cite{Arutyunov:2010gu} and
\cite{deLeeuw:2011rw}, focusing on non-supersymmetric orbifolds which are special cases of the 
general treatment of \cite{Beisert:2005he}. Here, we consider the construction of the Y-system for all 
cases considered in \cite{Beisert:2005he} and follow the approach of \cite{GKV,Gromov:2010dy},
thus bypassing the many subtleties of the rigorous TBA treatment or the S-matrix approach.
This makes our proposal an educated  conjecture which could be hopefully helpful for a more solid investigation.
The main point is that the orbifold projection breaks supersymmetry by boundary conditions on the string side.
This means that the powerful symmetry constraints of the untwisted case are inherited in the bulk. 
In more simple terms, the Y-system can be conjectured to be associated with the same Hirota dynamics where
the twisting parameters enter as rigid deformation parameters of polynomial solutions.

\medskip
In the non-supersymmetric case~\footnote{
We remark that, due to condensation of closed string tachyons, the orbifold symmetry can be spontaneously broken in 
 non-supersymmetric orbifold theories~\cite{Tseytlin:1999ii,ssb}.
}, 
we agree
with the results of \cite{Arutyunov:2010gu,deLeeuw:2011rw} and present a reciprocity respecting 
closed formula for the leading wrapping correction to twist-3 
operators in the orbifoldized $\mathfrak{sl}(2)$ subsector. For other orbifolds,
we illustrate various  examples where the calculation of leading wrapping corrections to physical states 
is definitely feasible as well as rather simple.

\bigskip
The plan of the paper is the following. In Sec.(\ref{sec:review}), we review the  
main features of orbifolds of AdS/CFT duality.  In Sec.(\ref{sec:ABA}), we present the asymptotic Bethe Ansatz
for orbifolds.  In Sec.(\ref{sec:Y-system-undeformed}), we illustrate the proposed twisted Y-system.
Finally,  in Sec.(\ref{sec:applications}) , we report various applications and checks.

\section{Orbifolds of AdS/CFT duality}
\label{sec:review}

In  orbifoldized AdS/CFT duality, we start from $\ads$ and quotient $S^{5}$ by a discrete 
isometry subgroup $\Gamma$ of $SO(6)$, or more generally of the spin cover $SU(4)$ as mentioned in the introduction. 
On the gauge theory side, we deal with what is called an orbifold projection (see for instance \cite{Lawrence:1998ja}). The initial $N$ D3-branes have $|\Gamma|\,N$ images. Thus, it is convenient to split the gauge index of $U(|\Gamma|\,N)$ as the pair $ (i, h)$, with $i=1, \dots, N$, and $h\in\Gamma$.
The action of the orbifold group is simply
$
g: (i, h)\mapsto (i, g\,h).
$
For example, if $\Gamma=\Z_{S}$, the index $I=1, \dots, SN$ is split into $S$ blocks of length $N$ and the action of $\Z_{S}$ is simply addition modulo $S$.
The Lagrangian of the orbifold projection is the same as in $\N=4$, with projected 
(under $P_{\Gamma} = \frac{1}{|\Gamma|}\sum_{g\in\Gamma} g$)
vector and chiral superfields .The group action on gauge indices spans the {\em regular} representation $\rho_{\rm reg}$. The vector and chiral superfields transform as 
\be
\mathcal{A}\sim \rho_{\rm reg}^{\oplus N}\otimes \overline\rho_{\rm reg}^{\oplus N}, \qquad
\Phi\sim \rho_{R}\otimes \rho_{\rm reg}^{\oplus N}\otimes \overline\rho_{\rm reg}^{\oplus N},
\ee
where $\rho_{R}$ is a 3-dimensional representation acting on the chiral field index $a=1,2,3$. 

\bigskip
After projection, it is easy to identify the invariant gauge fields and the associated orbifold gauge group. To this aim, 
one decomposes the regular representation  as 
\be
\rho_{\rm reg} = \bigoplus_{\lambda}\rho_{\lambda}^{\oplus N_{\lambda}},\qquad N_{\lambda}=\dim\rho_{\lambda},
\ee
where $\rho_{\lambda}$ are the irreducible representations of $\Gamma$.
From the invariance condition 
$
\mathcal{A} = \rho_{\rm reg}(g) \,\mathcal{A}\,\rho_{\rm reg}^{-1}(g) ,
$
we deduce that also $\mathcal{A}$ is block diagonal and, as a consequence, the orbifold gauge group is 
$
G_{\rm orb} = \bigotimes_{\lambda} U(NN_{\lambda}).
$
In other words, we simply  isolate the singlet representation in 
\be
\rho_{\rm reg}^{\oplus N}\otimes \overline\rho_{\rm reg}^{\oplus N} = \bigoplus_{\lambda, \lambda'}
\rho_{\lambda}\otimes \overline\rho_{\lambda'}\otimes \mathbb{C}^{NN_{\lambda}}\otimes
(\mathbb{C}^{*})^{NN_{\lambda'}} = \left[
\bigoplus_{\lambda}
\rho_{0}\otimes \mathbb{C}^{NN_{\lambda}}\otimes
(\mathbb{C}^{*})^{NN_{\lambda}}\right]\oplus\cdots.
\ee
The complexified coupling $\tau = \frac{4\pi i}{g^{2}}+\frac{\theta}{2\pi}$ breaks down to the 
 couplings 
 $
\tau_{\lambda} = \frac{N_{\lambda}}{|\Gamma|}\,\tau,
$
associated with the factors $U(NN_{\lambda})$.

\bigskip
The invariant matter can be computed in a similar way.
We have initially Weyl fermions $\Psi^{\alpha}_{IJ}$ in the adjoint of the gauge group with $\alpha$ in the $\mathbf{4}$ of $SU(4)$, and scalars $\Phi^{m}_{IJ}$ in the adjoint of the gauge group with $m$ in the $\mathbf{6}$ of $SO(6)$ which is $(\mathbf{4}\otimes\mathbf{4})_{\rm A}$ of $SU(4)$. For any representation $R$ we  define
the branching coefficients $a^{R}_{\lambda\lambda'}$ from 
$
\rho_{R}\otimes \rho_{\lambda} = \bigoplus_{\lambda'} a^{R}_{\lambda\lambda'}\,\rho_{\lambda'}.
$
Then, the singlets surviving projections are obtained from the relation
\be
\rho_{R}\otimes \rho_{\rm reg}^{\oplus N}\otimes \overline\rho_{\rm reg}^{\oplus N} = \left[\bigoplus_{\lambda, \lambda'}
a^{R}_{\lambda\lambda'}\,\rho_{0}\otimes\mathbb{C}^{NN_{\lambda}}
\otimes (\mathbb{C}^{*})^{NN_{\lambda'}}\right]\oplus\cdots.
\ee
Thus, we have $a^{R}_{\lambda\lambda'}$ fields transforming in the $(NN_{\lambda}, \overline{NN_{\lambda'}})$ bifundamental representation of the subgroup pair
$
U(NN_{\lambda})\times U(NN_{\lambda'})\subset G_{\rm orb}.
$
This information can be encoded in a {\em quiver diagram} \cite{Douglas:1996sw} where each representation $\rho_{\lambda}$ is associated with a node, and we draw $a^{R}_{\lambda\lambda'}$ directed arrows for each associated fields. It can be shown that there is a Yukawa coupling for all $(\mbox{fermion})^{2}\cdot\mbox{boson}$ triangles, and a quartic coupling for all $(\mbox{\rm boson})^{4}$ squares.

\bigskip
The residual supersymmetry is also easily identified.
Let $SU(2)\subset SU(3)\subset SU(4)$ be the standard embedding. It can be shown that if $\Gamma\subset SU(3)$,
then the residual supersymmetry of the orbifold projected theory is $\N=1$. If $\Gamma\subset SU(2)$, 
then it is $\N=2$. The multiplet identification goes as follows.
In the $\N=1$ case, we can reduce the $\bf{4}$ of $SU(4)$ under $SU(3)$ as 
${\bf 4} = {\bf 3}\oplus {\bf 1}$.
The ${\bf 3}$ is a triplet of  Weyl spinors. The singlet is the gaugino. The scalars organize with respect to $SU(3)$ 
as
\be
6  = [(3 \oplus 1)\otimes (3\oplus 1)]_{\rm A} = 
(3\otimes 3)_{\rm A}\oplus (3\otimes 1 + 1\otimes 3)_{\rm A} = \overline 3 \oplus 3.
\ee
Thus, we have a triplet and an antitriplet of real scalars. With the Weyl fermions, they build up three chiral superfields, as is known to happen when reducing the $\N=4$ hypermultiplet to $\N=1$. The gaugino pairs with the gauge potential  build a $\N=1$ vector superfield.

In the $\N=2$ case, we can reduce the ${\bf 4}$ of $SU(4)$ under $SU(2)$ as
$
{\bf 4} = {\bf 2}\oplus { 1}\oplus {\bf 1'}.
$
The ${\bf 2}$ is a doublet of  Weyl spinors. The singlets are two gauginos. The scalars organize with respect to $SU(2)$ 
as
\ba
6  &=& [(2 \oplus 1\oplus 1')\otimes (2\oplus 1\oplus 1')]_{\rm A} = 
(2\otimes 2)_{\rm A}\oplus 2(2\otimes 1 + 1\otimes 2)_{\rm A}\oplus (1\otimes 1')_{\rm A} = \nonumber \\
&=& 1\oplus 2\cdot {\bf 2}\oplus 1'.
\ea
Thus, we have two doublets of real scalars that combine with the doublet of Weyl fermions to build two chiral superfields. One gaugino pairs with the gauge fields and the other with the remaining two scalars in an additional $SU(2)$-invariant chiral superfield. This is indeed a $\N=2$ hypermultiplet.

\section{Asymptotic twisted Bethe Ansatz}
\label{sec:ABA}

The asymptotic Bethe Ansatz for  $\Z_{S}$ orbifold of planar \sym has been derived in \cite{Beisert:2005he}.
Here, we briefly review its main features as a preparation for the construction of the Y-system.

The group $\Z_{S}$ has $S$ (one-dimensional) 
irreducible representations where the cyclic element $g$ such that $g^{S}=1$ is
represented as 
\be
\rho_{n}(g) = \omega^{n},\qquad n=0, \dots, S-1, \qquad \omega=e^{\frac{2\pi\,i}{S}\,n}.
\ee 
The orbifold is thus a quiver theory with $S$ nodes and gauge group $U(N)^{\otimes S}$.
The product of representations is of course simply
\be
\rho_{n}\otimes \rho_{n'} = \rho_{n+n' \mod S}
\ee
In general, we start by representing $g$ on the Weyl fermions, {\em i.e.} on the {\bf 4} of $SU(4)$. This will take the form 
\be
g_{\bf 4} = \mbox{diag}(\omega^{n_{i}}), \qquad i=1, \dots, 4, \qquad n_{1}+\cdots+n_{4} = 0 \mod S.
\ee
The representation of $g$ on the {\bf 6} of $SU(4)$ follows from ${\bf 6}=({\bf 4}\times {\bf 4})_{\rm A}$ .
Thus
\be
{\bf 4} = \bigoplus_{i=1}^{4} \rho_{n_{i}}, \qquad
{\bf 6} = \mathop{\bigoplus_{i=1}^{4}}_{i<j} \rho_{n_{i}+n_{j}}.
\ee
This allows to compute $a^{\bf 4}_{nm}$ and $a^{\bf 6}_{nm}$ and hence the quiver diagram. 
The amount of residual supersymmetry can be easily recognized as we mentioned in the previous section.
This is a somewhat important issue since the leading order at which wrapping effects appear is delayed in perturbation theory by supersymmetry, as it happens in the unorbifoldized $\ads$ case.

We parametrize  the cyclic element $g_{\bf 4}$ in the fundamental of $SU(4)$ 
acting on Weyl fermions as
\be
g_{\bf 4} = \mbox{diag}(\omega^{-t_{1}}, \omega^{t_{1}-t_{2}},
\omega^{t_{2}-t_{3}}, \omega^{t_{3}}), \qquad  \omega=e^{2\pi i/S},
\ee
where $S=2, 3, \dots$ determines the orbifold order.
Fields with definite $SU(4)$ charges obey 
\be
X = \omega^{s_{X}}\,\rho\,X\,\rho^{-1},
\ee
where $\rho$ is the representation of the orbifold action on the gauge indices. This relation can be used to bring all 
twist matrices together. We are led to consider the operators
\be
\mathcal{O} = \mbox{Tr}(\rho^{T}\,\mathcal{W}_{A_{1}}\cdots\mathcal{W}_{A_{L}}), 
\qquad \mathcal{W}\in\{D^{n}\varphi, D^{n}\psi, D^{n}F\}
\ee
where $T=0, \dots, S-1$ labels the twist sector and, for long enough operators, is a conserved quantum number. These operators are non (trivially) vanishing if the sum of twist charges vanishes
\be
\sum_{i=1}^{L} s_{A_{i}} = 0 \mod S.
\ee

The asymptotic one-loop Bethe equations, cyclicity condition and twist constraint turn out to be (here 
$J=7$ and $L$ is the length of the Bethe states)
\ba
&& 
\omega^{T\,s_{j}}\,\left(\frac{u_{j,k}-\frac{i}{2}V_{j}}{u_{j,k}+\frac{i}{2}V_{j}}\right)^{L}\,
\mathop{\prod_{j'=1}^{J}\prod_{k'=1}^{K_{j'}}}_{(j,k)\neq(j',k')} 
\frac{u_{j,k}-u_{j',k'}+\frac{i}{2}M_{jj'}}{u_{j,k}-u_{j',k'}-\frac{i}{2}M_{jj'}} = 1, \\
&& 
\omega^{T\,s_{0}}\,\prod_{j=1}^{J}\prod_{k=1}^{K_{j}} 
\frac{u_{j,k}+\frac{i}{2}V_{j}}{u_{j,k}-\frac{i}{2}V_{j}}=1, \\
&& 
\omega^{-L\,s_{0}}\,\prod_{j=1}^{J}\omega^{-K_{j}\,s_{j}} = 1.
\ea
Here, $M$ is the Cartan matrix of the $\mathfrak{psu}(2,2|4)$ algebra, $V_{j}$ are the weights determining the 
transformation rules of the associated spin chain, and $\{s_{j}\}_{j=0, \dots, 7}$ are twist exponents. Their values 
for $j>0$ can be computed starting from the oscillator representation of the Dynkin diagram. The value of $s_{0}$ is computed from the vacuum state of the specific Dynkin diagram.
The explicit values for the {\em higher} form of the Dynkin diagram \cite{Beisert:2005fw} are shown in Fig.~(\ref{fig:higher-charges}).
\begin{figure}[h]
\begin{center}
\begin{minipage}{300pt}
\setlength{\unitlength}{1pt}%
\small\thicklines%
\begin{picture}(300,55)(-50,-10)
\put(-40,00){\circle{15}}%
\put(-40,15){\makebox(0,0)[b]{$-t_2$}}%
\put(  0,00){\circle{15}}%
\put(  0,15){\makebox(0,0)[b]{$t_1$}}%
\put(  7,00){\line(1,0){26}}%
\put( 40,00){\circle{15}}%
\put( 40,00){\makebox(0,0){$-$}}%
\put( 40,15){\makebox(0,0)[b]{$0$}}%
\put( 47,00){\line(1,0){26}}%
\put( 80,00){\circle{15}}%
\put( 80,30){\makebox(0,0)[b]{$t_1-t_2$}}%
\put( 87,00){\line(1,0){26}}%
\put(120,00){\circle{15}}%
\put(120,00){\makebox(0,0){$+$}}%
\put(120,15){\makebox(0,0)[b]{$2t_2-t_1-t_3$}}%
\put(127,00){\line(1,0){26}}%
\put(160,00){\circle{15}}%
\put(160,30){\makebox(0,0)[b]{$t_3-t_2$}}%
\put(167,00){\line(1,0){26}}%
\put(200,00){\circle{15}}%
\put(200,00){\makebox(0,0){$-$}}%
\put(200,15){\makebox(0,0)[b]{$0$}}%
\put(207,00){\line(1,0){26}}%
\put(240,00){\circle{15}}%
\put(240,15){\makebox(0,0)[b]{$t_3$}}%
\put( -5,-5){\line(1, 1){10}}%
\put( -5, 5){\line(1,-1){10}}%
\put( 75,-5){\line(1, 1){10}}%
\put( 75, 5){\line(1,-1){10}}%
\put(155,-5){\line(1, 1){10}}%
\put(155, 5){\line(1,-1){10}}%
\put(235,-5){\line(1, 1){10}}%
\put(235, 5){\line(1,-1){10}}%
\end{picture}
\end{minipage}\vspace{0.5cm}
 \caption{Values of the twist exponents $\{s_{j}\}$ for the higher Dynkin diagram.}
 \label{fig:higher-charges}
 \end{center}
\end{figure}
For this Dynkin diagram we have the following Cartan matrix and weights
\be
M = {\small \left(
\begin{array}{ccccccc}
 & 1 \\
 1 & -2 & 1 \\
 & 1 & & -1 \\
 && -1 & 2 & -1 \\
 &&& -1 && 1 \\
 &&&& 1 & -2 & 1 \\
 &&&&& 1 
\end{array}
\right)},\quad V_{j} = \delta_{j,4}.
\ee
As discussed in \cite{Beisert:2005he}, the twist exponents do not change when the Bethe equations are replaced by 
their all-order asymptotic form, including dressing corrections. We shall not need them in explicit form in the 
following manipulations which can be done at one-loop.

\subsection{$\mathfrak{sl}(2)$ grading}

In the following, it will be convenient to dualize the fermionic nodes in order to write the Bethe equations 
in $\mathfrak{sl}(2)$ grading. To this aim, and following standard manipulations, we 
introduce $\widetilde K_{j}$ dual roots at fermionic nodes and associated polynomials 
$Q_{j} = \prod_{k=1}^{K_{j}}(u-u_{j,k})$, $\widetilde Q_{j} = \prod_{k=1}^{\widetilde 
K_{j}}(u-\widetilde u_{j,k})$, according to 
\ba
\omega^{T\,s_{1}}\,Q_{2}^{+}-Q_{2}^{-} = (\omega^{T\,s_{1}}-1)\,Q_{1}\,\widetilde Q_{1}, && \qquad K_{1}+\widetilde K_{1} = K_{2}, \nonumber , \\
\omega^{T\,s_{3}}\,Q_{2}^{+}Q_{4}^{-}-Q_{2}^{-}Q_{4}^{+} = (\omega^{T\,s_{3}}-1)\,Q_{3}\,\widetilde Q_{3}, && \qquad K_{3}+\widetilde K_{3} = K_{2}+K_{4}, \nonumber \\
\omega^{T\,s_{5}}\,Q_{4}^{-}Q_{6}^{+}-Q_{4}^{+}Q_{6}^{-} = (\omega^{T\,s_{5}}-1)\,Q_{5}\,\widetilde Q_{5}, && \qquad K_{5}+\widetilde K_{5} = K_{4}+K_{6}, \nonumber  \\
\omega^{T\,s_{7}}\,Q_{6}^{+}-Q_{6}^{-} = (\omega^{T\,s_{7}}-1)\,Q_{7}\,\widetilde Q_{7}, && \qquad K_{7}+\widetilde K_{7} = K_{6}.
\ea
Again, we stress that these are one-loop relations with simple extension to the all-order deformation.
Replacing in the Bethe equations we can rewrite them in the $\mathfrak{sl}(2)$ form as 
 \be 
 \label{eq:sl2bethe}
\omega^{T\,\widetilde s_{j}}\,\left(\frac{u_{j,k}-\frac{i}{2}V_{j}}{u_{j,k}+\frac{i}{2}V_{j}}\right)^{L}\,
\mathop{\prod_{j'=1}^{J}\prod_{k'=1}^{K_{j'}}}_{(j,k)\neq(j',k')} 
\frac{u_{j,k}-u_{j',k'}-\frac{i}{2}M_{jj'}}{u_{j,k}-u_{j',k'}+\frac{i}{2}M_{jj'}} = 1, 
\ee 
where the dualized twist exponents are 
\be
\begin{array}{|c|c|c|c|c|c|c|}
\hline
 \widetilde s_{1} & \widetilde s_{2} & \widetilde s_{3} & \widetilde s_{4} & \widetilde s_{5} & \widetilde s_{6} & \widetilde s_{7} \\
 -t_{1} & 2t_{1}-t_{2} & t_{2}-t_{1} & 0 & t_{2}-t_{3} & 2t_{3}-t_{2} & -t_{3}\\
\hline
\end{array}
\ee
The (unaltered) twisted ciclicity and twist constraints read 
\ba
\label{eq:cyc}
&& \omega^{T\,s_{0}}\,\prod_{k=1}^{K_{4}} 
\frac{u_{4,k}+\frac{i}{2}}{u_{4,k}-\frac{i}{2}}=1, \\
&& 
L\,s_{0}+\sum_{j=1}^{7} \widetilde K_{j}\,\widetilde s_{j} = 0 \mod S.
\ea
We remark that $s_{0}$ does not depend on grading, and in the twist constraint, we understand 
$K_{j}=\widetilde K_{j}$ for the bosonic nodes.

\section{Twisted Y-system for $\Z_{S}$ orbifolds}
\label{sec:Y-system-undeformed}

The  spectrum of relativistic 1+1 dimensional integrable theories has been suggested~\cite{Zamolodchikov:1991et} 
to be captured by the universal set of functional quadratic Hirota equations taking the form
\be
\label{Tsystem}
  T_{a,s}(u+i/2)\,T_{a,s}(u-i/2) =T_{a+1,s}(u)\,T_{a-1,s}(u)+T_{a,s+1}(u)\,T_{a,s-1}(u)\;.
\ee
For the $\ads$ superstring theory 
 the system of Hirota equations
should be the same, with the functions $T_{a,s}(u)$ being non-zero
only inside the infinite T-shaped domain of the ${a,s}$ integer lattice, shown in Fig.~(\ref{Fig:figysys}).
\FIGURE[ht]
{\label{Fig:figysys}
    \begin{tabular}{cc}
    \includegraphics[scale=0.4]{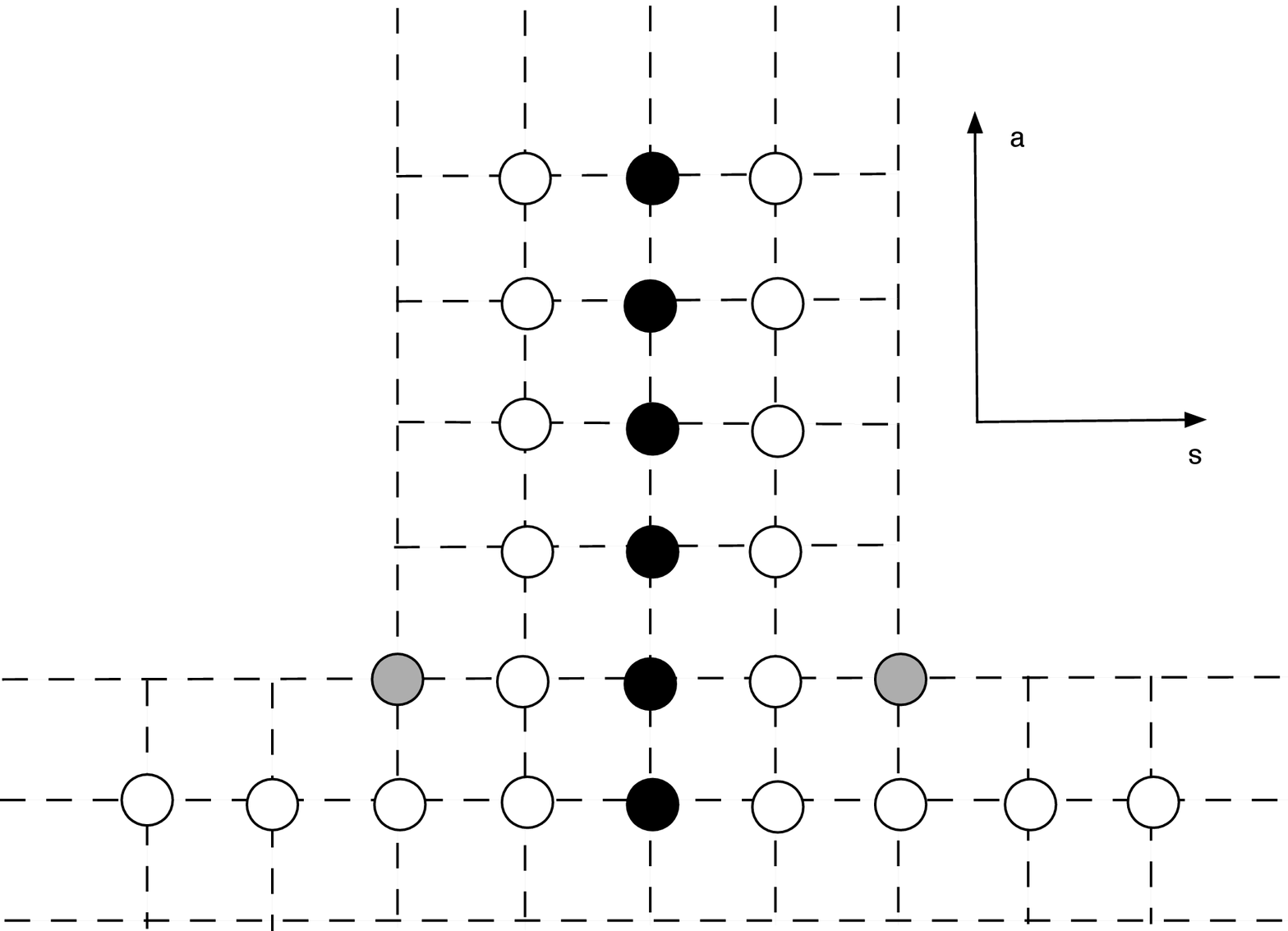}
    \end{tabular}
    \caption{Graphical representation of the Y-system and T-system.
    Circles
    correspond to Y-functions. Intersections of grid lines in the T-hook correspond to T-functions.}
}
Physical quantities can be computed by introducing the gauge invariant $Y$-functions
\begin{equation}
\label{YT}
Y_{a,s}=\frac{T_{a,s+1}T_{a,s-1}}{T_{a+1,s}T_{a-1,s}}\;\; ,
\end{equation}
which obey another set of functional equations called the Y-system:\begin{equation}
\label{eq:Ysystem}
\frac{Y_{a,s}^+ Y_{a,s}^-}{Y_{a+1,s}Y_{a-1,s}}
  =\frac{(1+Y_{a,s+1})(1+Y_{a,s-1})}{(1+Y_{a+1,s})(1+Y_{a-1,s})} \, ,
\end{equation}
The T- and Y-systems should additionally be supplemented with a particular set of analytical properties imposed on T- and Y-functions. The deep analysis of these  properties in the AdS/CFT case is discussed in \cite{TBA1}.

\medskip
The energy and momentum of magnon excitations in the theory are described
in terms of $x(u)$, defined by
    \be
    \label{xdef}
        x + \frac{1}{x} = \frac{u}{g},
    \ee
where the relation between the coupling $g$ and the 't Hooft coupling $\lambda$ is $\lambda = 16\pi^{2}g^{2}$.
The {\em mirror} and {\em physical} branches of this function are defined as
   \be
    \label{xBranches}
        x^{\rm ph}(u)=\frac{1}{2}\,\left(
        \frac{u}{g}+\sqrt{\frac{u}{g}-2}\;\sqrt{\frac{u}{g}+2}\right)\;\;,\;\;
        x^{\rm mir}(u)=\frac{1}{2}\left( \frac{u}{g}+i\sqrt{4-\frac{u^2}{g^2}}\right) \,.
    \ee
    The energy and momentum of a bound state with $n$
    magnons are~\footnote{We use the by now standard notation 
 \be
	f^\pm \equiv f(u\pm i/2),\  f^{[+a]} \equiv f(u + ia/2).
\ee
}
 \begin{equation}
\label{epsandp}
\varepsilon_n(u)= n+\frac{2ig}{x^{[+n]}}-\frac{2ig}{x^{[-n]}}\;\;,\;\;
p_n(u)=\frac{1}{i}\log\frac{x^{[+n]}}{x^{[-n]}}\;.
\end{equation}
Finally, the exact energy of a state is given by
\be
\label{Efull}
E=\sum_{j}\varepsilon_1^{\rm ph} (u_{4,j})+\delta E\;\;,\;\;
\delta E = \sum_{a=1}^\infty \int_{-\infty}^{\infty}\frac{du}{2\pi i}
\,\,\frac{\partial\varepsilon_a^{*} (u)}{\partial u} \log\left(1+Y_{a,0}^* (u)\right)\;,
\ee
where the rapidities $u_{4,j}$ are fixed by the exact Bethe ansatz equations
\be
Y_{1,0}^{\rm ph} (u_{4,j})=-1.
\ee
Here, $Y_{a,0}^*$ denotes the function $Y_{a,0}$ evaluated at mirror kinematics.

\subsection{Twisting and matching}

The clever idea of \cite{GKV} is that, for asymptotically large size $L$, 
it is possible to solve the $Y$-system explicitly since the massive nodes 
$Y_{a,0}$ decouple and the Y-system splits into two wings $\mathfrak{su}_{\rm L}(2|2)\oplus \mathfrak{su}_{\rm R}(2|2)$. At weak coupling, the solution found in this way can also be used to compute leading wrapping corrections 
at fixed finite $L$.

The explicit form of the transfer matrices can be derived from the solutions of the Hirota equation for a domain called L-hook (one half of the T-hook diagram in Fig.(\ref{Fig:figysys})) and written in the form of a generating 
functional~\cite{hirota}. The construction of \cite{twisted-hirota} allows for certain free parameters that have been exploited in \cite{Gromov:2010dy}
 to build the transfer matrices for the $\beta$-deformed theory. Here, we shall exploit them to compute the orbifold 
 transfer matrices. 
Let us introduce the quantities
\be
\widetilde R^{(\pm)} =  \prod_{j=1}^{K_{4}}\frac{x(u)-x^{\mp}_{4,j}}{\sqrt{x^{\mp}_{4,j}/u^{(0)\,\mp}_{4,j}}}, \qquad
\widetilde B^{(\pm)} =  \prod_{j=1}^{K_{4}}\frac{\frac{1}{x(u)}-x^{\mp}_{4,j}}{\sqrt{x^{\mp}_{4,j}/u^{(0)\,\mp}_{4,j}}},
\ee
where $u_{4,j} = u_{4,j}^{(0)}+\mathcal{O}(g^{2})$, and $x_{4,j}^{\pm} = x^{\pm}(u_{4,j})$.
The middle node Y-functions for large $L$ can then be written as
\be
\label{Yaasymp}
Y_{a,0}\simeq \left(\frac{x^{[-a]}}{x^{[+a]}}\right)^L \,T_{a,1}^{\ell}\,T_{a,1}^{r}\,
\Phi_a\;,
\ee
where the (simplified $\mathfrak{sl}(2)$) fused scalar factor is
\ba
\label{eq:twistedfusion}
\Phi_a(u) &=& \prod_{n=-\frac{a-1}{2}}^{\frac{a-1}{2}}\Phi(u+i\,n), \nonumber \\
\Phi(u) &=& \frac{1}{\lambda_{0}}
\frac{\widetilde B^{(+)+}}{\widetilde B^{(-)-}}\frac{\widetilde R^{(-)-}}{\widetilde R^{(+)+}}\prod_{j=1}^{K_4}\sigma^2(u,u_{4,j}).
\ea
Here, $\sigma (u,v)$ is the dressing factor, and $\lambda_{0}$ is given by ~\footnote{Notice that since 
the right hand side of (\ref{eq:lambda0}) does not depend on $g$, it also implies the following condition on the one-loop Bethe roots
\be
\prod_{j=1}^{K_{4}}\frac{u_{4,j}^{(0)}-\frac{i}{2}}{u_{4,j}^{(0)}+\frac{i}{2}} = \omega^{-T\,t_{2}}.
\ee 
}
\be
\label{eq:lambda0}
\lambda_{0} = \prod_{j=1}^{K_{4}}\frac{x^{-}_{4,j}}{x^{+}_{4,j}} =  \omega^{-T\,t_{2}},
\ee 
and comes from the following weak coupling expansion
\be
\frac{\widetilde B^{(+)+}}{\widetilde B^{(-)+}} = \lambda_{0}+\mathcal{O}(g^{2}).
\ee
The leading order  wrapping correction can be found from \eq{Efull} and reads:
\be
\label{eq:wrap-leading}
\delta E^{\rm LO}\simeq -\sum_{a=1}^{\infty}\int_{-\infty}^{\infty}\frac{du}{\pi} Y_{a,0}^*(u),
\ee
where values of the Bethe roots which enter the expression for $Y_{a,0}$ should be obtained from the asymptotic 
 twisted Bethe equations. 

\medskip
The twisted transfer matrices (building the $Y$-functions) are obtained from the following twisted generating series
(with a similar one for the right wing with $\ell\to r$ and a relabeling of simple roots)
\ba
\label{eq:twistedgenerating}
\lefteqn{\sum_{a=0}^{\infty} (-1)^{a}\,T_{a,1}^{\rm\ell}\left(u+i\frac{1-a}{2}\right)\,\ddb^{a} = } && \\
&& 
\left(1-\lambda_{1}^{\ell}\,\frac{Q_{3}^{+}}{Q_{3}^{-}}\,\ddb\right)^{-1}\,
\left(1-\lambda_{2}^{\ell}\,\frac{Q_{3}^{+}}{Q_{3}^{-}}\frac{Q_{2}^{--}}{Q_{2}}\frac{\widetilde R^{(+)-}}
{\widetilde R^{(-)-}}\,\ddb\right)\times \nonumber \\
&& \left(1-\lambda_{3}^{\ell}\,\frac{Q_{2}^{++}}{Q_{2}}\frac{Q_{1}^{-}}{Q_{1}^{+}}\frac{\widetilde R^{(+)-}}{\widetilde R^{(-)-}}
\,\ddb\right)\,
\left(1-\lambda_{4}^{\ell}\,\frac{Q^{-}_{1}}{Q_{1}^{+}}\frac{\widetilde B^{(+)+}}{\widetilde B^{(-)+}}\frac{\widetilde R^{(+)-}}{\widetilde R^{(-)-}}\ddb\right)^{-1},
\nonumber
\ea
where where $\ddb = e^{-i\partial_{u}}$ is the shift operator. For the right wing, we have a similar expression for the $T_{a,1}^{r}$ functions with subscripts of functions $Q_i$ changed according to $1,2,3\to 7,6,5$. 

The complex constants $\lambda^{\ell, r}_{1,2,3,4}$ are arbitrary and will be fixed by matching the all-order Bethe equations~\footnote{
Notice that the various factors $\lambda_{0}$ have the same role of the analogous momentum dependent
phases appearing in $AdS_{4}\times\mathbb{CP}^{3}$ duality where one has  to introduce them
\cite{Gromov:2009at} in order to match the Bethe equations of \cite{Gromov:2008qe}.
Anyhow, they could be absorbed in the definition of  $\lambda^{\ell, r}_{i}$. 
}. 
This can be done at one-loop since we are simply identifying them in terms of the rigid phases entering the 
Bethe Ansatz for the orbifold.

The above generating functional agrees with the known untwisted one for cyclic states with $\lambda_{0}=1$
and by setting $\lambda^{\ell, r}_{i}=1$. A comparison with other presentations, like that in \cite{Gromov:2010dy},
can be easily checked by using the results of App.~(\ref{app:gauge}).

\bigskip
Matching the one-loop Bethe equations associated with the twisted transfer matrices with (\ref{eq:sl2bethe})
we find immediately
\ba
\lambda_{3}^{\ell} &=& \lambda_{0}\,\lambda_{4}^{\ell}\,\omega^{-T\,\widetilde s_{1}}, \nonumber\\
\lambda_{2}^{\ell} &=& \lambda_{0}\,\lambda_{4}^{\ell}\,\omega^{-T\,(\widetilde s_{1}+\widetilde s_{2})}, \nonumber\\
\lambda_{1}^{\ell} &=& \lambda_{0}\,\lambda_{4}^\ell\,\omega^{-T\,(\widetilde s_{1}+\widetilde s_{2}+\widetilde s_{3})}, \nonumber\\
\lambda_{1}^{r} &=& \lambda_{0}^{-1}\,(\lambda_{4}^{\ell})^{-1}\,\omega^{T\,(\widetilde s_{1}+\widetilde s_{2}+\widetilde s_{3})}, \\
\lambda_{2}^{r} &=& \lambda_{0}^{-1}\,(\lambda_{4}^{\ell})^{-1}\,\omega^{T\,(\widetilde s_{1}+\widetilde s_{2}
+\widetilde s_{3}+\widetilde s_{5})}, \nonumber\\
\lambda_{3}^{r} &=& \lambda_{0}^{-1}\,(\lambda_{4}^{\ell})^{-1}\,\omega^{T\,(\widetilde s_{1}+\widetilde s_{2}
+\widetilde s_{3}+\widetilde s_{5}+\widetilde s_{6})}, \nonumber\\
\lambda_{4}^{r} &=& \lambda_{0}^{-2}\,(\lambda_{4}^{\ell})^{-1}\,\omega^{T\,(\widetilde s_{1}+\widetilde s_{2}
+\widetilde s_{3}+\widetilde s_{5}+\widetilde s_{6}+\widetilde s_{7})}.\nonumber
\ea
The factor $\lambda_{4}^{\ell}$ is a gauge parameter which cancels in the $Y$-functions. 
We choose $\lambda_{4}^{\ell} = \lambda_{0}^{-1}$ and evaluate the functional with 
\be
\label{eq:twistedlambda}
\begin{array}{ccc}
\lambda_{1}^{\ell} &=& 1\\
\lambda_{2}^{\ell} &=& \omega^{T\,(t_{2}-t_{1})} \\
\lambda_{3}^{\ell} &=& \omega^{T\,t_{1}} \\
\lambda_{4}^{\ell} &=& \omega^{T\,t_{2}}
\end{array}, \qquad\qquad
\begin{array}{ccc}
\lambda_{1}^{r} &=& 1 \\
\lambda_{2}^{r} &=& \omega^{T\,(t_{2}-t_{3})} \\
\lambda_{3}^{r} &=& \omega^{T\,t_{3}} \\
\lambda_{4}^{r} &=& \omega^{T\,t_{2}}
\end{array}. 
\ee
One can check that in all presented applications, the mirror quantities $Y_{a,0}^{*}$ are always real. The factor
$\lambda_{0}^{-a}$ coming from $\Phi$ in (\ref{eq:twistedfusion}) (and invisible for orbifolds with untwisted ciclicity condition)
 is crucial for this.

\section{Applications to the $\mathfrak{sl}(2)$ sector of various orbifolds}
\label{sec:applications}

\subsection{Non-supersymmetric $(t_{1}, t_{2}, t_{3}) = (t_{1},0,t_{3})$ orbifolds}

The only non-trivial $Q$ polynomial is $Q_{4}$. 
We consider $t_{2}=0$ and generic non zero $t_{1}$, $t_{3}$.
We automatically solve the cyclicity constraint (\ref{eq:cyc}) for states with an even $Q_{4}$ since $s_{0}=-t_{2}=0$ and the Bethe roots appear in opposite pairs~\footnote{Notice that for $t_{2}=0$ it is impossible to satisfy the 
ciclicity constraint with a single excitation.}. Under this assumption, 
the contributions to (\ref{Yaasymp}) {\em evaluated in the mirror dynamics} are

\medskip\noindent
\underline{\em Dispersion}

\medskip\noindent
This is the universal factor coming from $(x^{[-a]}/x^{[a]})^{L}$ evaluated in the mirror dynamics
\be
\left(\frac{4g^{2}}{a^{2}+4u^{2}}\right)^{L} \ee

\medskip\noindent
\underline{\em{twisted $su(2|2)$  wing} }

\medskip
\noindent
After a straightforward computations one finds the following compact efficient formula (which is valid
assuming that $Q_{4}$ is an even polynomial)
\be
\label{eq:wing}
T_{a,0}^{\ell\ \ *} = (-2+\lambda_{2}^{\ell}+\lambda_{3}^{\ell})^{}\,\frac{(-1)^{a+1}}{Q_{4}^{[1-a]}}
\mathop{\sum_{p=-a+1}^{a-1}}_{\Delta p = 2} Q_{4}^{[p]},
\ee
with a similar result for the right wing. Notice that, since $t_{2}=0$, we also have
\ba
-2+\lambda_{2}^{\ell}+\lambda_{3}^{\ell} &=& 4\,\sin^{2}\left(\frac{\pi\,T\,t_{1}}{S}\right), \\
-2+\lambda_{2}^{r}+\lambda_{3}^{r} &=& 4\,\sin^{2}\left(\frac{\pi\,T\,t_{3}}{S}\right).\nonumber
\ea

\medskip\noindent
\underline{\em Fusion of scalar factors}

\medskip\noindent
After some manipulations, one finds (again, for an even polynomial $Q_{4}$) the formula
\be
\Phi^{*}_{a} = [(Q^{+}_{4}(0)]^{2} \,\frac{Q_{4}^{[1-a]}}{Q_{4}^{[-1-a]}Q_{4}^{[a-1]}Q_{4}^{[a+1]}}.
\ee

\subsubsection{1-particle states}

One particle states are not physical since they cannot satisfy the ciclicity constraint. For a discussion of their
properties and of the finite size corrections to the dispersion relations see the detailed discussion in \cite{Arutyunov:2010gu}.

\subsubsection{2-particle states}

Two particle states are associated with roots $u_{4} = \{v, -v\}$ where the Bethe equations determine
$v = \frac{1}{2}\cot\left(
\frac{\pi\,n}{L+1}\right)$, with $n=1, \dots, L$. Taking $Q_{4}(u)=u^{2}-v^{2}$ and evaluating the finite sum in the second factor in (\ref{eq:wing})
we find
\be
\frac{(-1)^{a+1}}{Q_{4}^{[1-a]}}
\mathop{\sum_{p=-a+1}^{a-1}}_{\Delta p = 2} Q_{4}^{[p]} = (-1)^{a+1}\frac{a}{12}
\frac{12\,u^{2}-12\,v^{2}+1-a^{2}}{\left(u-i\,\frac{a-1}{2}\right)^{2}-v^{2}}.
\ee
This perfectly agrees with the twisted transfer matrices of \cite{Arutyunov:2010gu} after combining the left and
right wings. The calculation of wrapping is then straightforward. For instance, for $L=2$ and $v=\frac{1}{2\sqrt 3}$
we get
\be
\delta E^{\rm LO} = -\frac{16}{3}\,\sin^{2}\left(\frac{\pi T}{S}\,t_{1}\right)\,
\sin^{2}\left(\frac{\pi T}{S}\,t_{3}\right)\,g^{4}.
\ee
Dividing by $2^{2L}$ due to a factor 2 in the definition of the coupling and taking $t_{1}=t_{3}$ we 
fully agree with the first line of Tab.~(4) of \cite{Arutyunov:2010gu}. The other entries are recovered as well.

\bigskip
If $t_{1}$ or $t_{3}$ vanish, then one has additional supersymmetry and  the leading wrapping order is delayed. 
For instance~\footnote{Of course, the compact formula (\ref{eq:wing}) is not sufficient to derive this result and one can improve it or just expand the 
general transfer matrices.}, taking $t_{1}\neq 0$ and $t_{3}=0$, we find, again  for $L=2$ and $v=\frac{1}{2\sqrt 3}$, 
\be
\delta E^{\rm LO} = 24\,\sin^{2}\left(\frac{\pi T}{S}\,t_{1}\right)\,g^{6}.
\ee
Working out the correction for the other states, we fully agree with the $\beta$-deformed results in Tab.(3) of \cite{Arutyunov:2010gu} after dividing out by $2^{2L}$ 
and identifying the $\beta$ deformation parameter via $\beta = T/(SL)$. Indeed, this orbifold has one twisted wing
and one undeformed wing precisely as in the $\beta$-deformed theory.

\subsubsection{$N$-particle states for $L=2, 3$}

Since the Bethe equations are undeformed, we can exploit the known Baxter polynomials for $L=2, 3$
as functions of the number of magnons $N$. They are (see for instance \cite{Beccaria:2008pp} and references
therein )
\ba
\label{eq:baxter2}
L=2, && \qquad Q_{4}(u) = {}_{3}F_{2}\left(\begin{array}{c}
-N, \ N+1,\ \frac{1}{2}+i\,u \\
1,\ 1
\end{array}\, ; 1\right), \\
L=3, && \qquad Q_{4}(u) = {}_{4}F_{3}\left(\begin{array}{c}
-\frac{N}{2}, \ \frac{N}{2}+1,\ \frac{1}{2}+i\,u,\ \frac{1}{2}-i\,u \\
1,\ 1, \ 1
\end{array}\, ; 1\right).
\ea
Inserting them in the $Y$-system equations we obtain  explicit formulae for the leading wrapping 
correction. 
For $L=2$, we find the simple result
\be
\label{eq:L2wrap}
\delta E^{\rm LO}_{L=2} = -\frac{32}{N\,(N+1)}\,g^{4}\,\sin^{4}(\pi\,T/S).
\ee
This is in perfect agreement with the results of \cite{deLeeuw:2011rw}.
For $L=3$, we find (in this case $N$ must be even)~\footnote{As usual
\be
S_{a}(n) = \sum_{p=1}^{n} (\mbox{\scriptsize sign(a)})^{p}\,p^{-|a|},\qquad 
S_{a, b, \dots}(n) = \sum_{p=1}^{n} (\mbox{\scriptsize sign(a)})^{p}\,p^{-|a|}\,S_{b, \dots}(p). 
\ee
}
\be
\label{eq:L3wrap}
\delta E^{\rm LO}_{L=3} = -\frac{32}{(N+1)^{2}}\left[
S_{3}\left(\frac{N}{2}\right)+6\,\zeta_{3}
\right]\,g^{6}\,\sin^{4}(\pi\,T/S).
\ee

\bigskip
{\bf Remark:} These expressions are reciprocity respecting in the sense of, say, \cite{Beccaria:2010tb}.
In other words, the large $N$ expansion of (\ref{eq:L2wrap}) is in {\em integer} inverse powers of $J^{2}=N(N+1)$ (this is trivial) and the large $N$ expansion of (\ref{eq:L3wrap}) is in {\em integer} inverse powers of $J^{2}=\frac{N}{2}(\frac{N}{2}+1)$ as one can easily check 
\ba
\delta E^{\rm LO}_{L=3} &=& -\frac{56 \zeta _3}{J^2}+\frac{14 \zeta _3+4}{\left(J^2\right)^2}+\frac{-\frac{7 \zeta _3}{2}-3}{\left(J^2\right)^3}+\frac{\frac{7 \zeta _3}{8}+\frac{25}{12}}{\left(J^2\right)^4}+\frac{-\frac{7 \zeta
   _3}{32}-\frac{89}{48}}{\left(J^2\right)^5}+\cdots. 
   \ea
Notice also that both $\delta E^{\rm LO}_{L=2,3}$ are suppressed as $1/N^{2}$ at large spin $N$.

\bigskip

\subsection{$\N=2$ supersymmetric orbifolds $(t_{1},t_{2},t_{3})=(0,1,0)$    }

Let us consider   $(t_{1}, t_{2}, t_{3})=(0,1,0)$ which is $\N=2$ supersymmetric.  The $\N=2$ supersymmetry shows up in the fact that the $T$ functions vanish at naive leading order. We thus get an additional $g^{2}$ factor from each wing and the leading wrapping correction is of order
$g^{2L+4}$ instead of the naive $g^{2L}$.

\subsubsection{1-particle states}

Let us  consider one-excitation states. For a $\Z_{S}$ orbifold, the unique central node root $v$ must obey the Bethe equation, ciclicity and twist constraints ($\omega=e^{\frac{2\pi\,i}{S}}$)
\be
\frac{x^{+}(v)}{x^{-}(v)}=\omega^{T}, \qquad
\left[\frac{x^{+}(v)}{x^{-}(v)}\right]^{L}=1, \qquad
-L = 0 \mod S.
\ee
A minimal solution is obtained with $L=S$, $T=1$, and reads
\be
v = \frac{1}{2}\cot\left(\frac{\pi}{S}\right)+2\,\sin\left(\frac{2\pi}{S}\right)\,g^{2}+\cdots~.
\ee
The Y-functions can be computed for general $S$ and turn out to be
\ba
Y_{a,0}^{*\ \rm LO}(u) &=& g^{4+2S}\,\frac{4^{S+5}\,a^{2}}{(a^{2}+4\,u^{2})^{S+2}}\,\frac{\left[
a^{2}+4\,u^{2}-\csc^{2}\left(\frac{\pi}{S}\right)\right]^{2}\,\sin^{8}\left(\frac{\pi}{S}\right)
}{f_{a}(u)\,f_{-a}(u)}, \\
f_{a}(u) &=& 
-2-2a-a^{2}-4u^{2}+(2a+a^{2}+4u^{2})\cos\left(\frac{2\pi}{S}\right)+4\,u\,\sin\left(\frac{2\pi}{S}\right),
\ea
The wrapping correction is evaluated as usual from Eq.~ (\ref{eq:wrap-leading}). The results for the first cases are
%
%
%
%
%
%
%
\ba
\label{eq:spin1}
\delta E^{\rm LO}_{S=2} &=&128\, (4\, \zeta_{3}-5\, \zeta_{5}) \,g^{8}, \\
\delta E^{\rm LO}_{S=3} &=& 180\, \left(6\, \zeta _5-7\, \zeta _7\right)\,g^{10}, \\
\delta E^{\rm LO}_{S=4} &=& -32 \left(8\,\zeta _5-56\, \zeta _7+63\, \zeta _9\right)\,g^{12}.
\ea
These can be checked to be in agreement with the one-particle wrapping corrections computed in \cite{Arutyunov:2010gu} (see their Eqs.~(6.2, 6.3) for model II).

\subsubsection{A simple class of 2-particle states}

In a $\Z_{S}$ orbifold, the 2-particle cyclicity constraint and Bethe equations for two rapidities $u$, $v$,
read
\ba
&&\frac{x^{-}(u)}{x^{+}(u)}
\frac{x^{-}(v)}{x^{+}(v)}
 = \lambda_{0} = e^{-\frac{2\pi\,i}{S}}, \\
&&\left(\frac{x^{-}(u)}{x^{+}(u)}\right)^{L}\,
\frac{x^{-}(u)-x^{+}(v)}{x^{+}(u)-x^{-}(v)}\frac{1-1/x^{+}(u)x^{-}(v)}{1-1/x^{-}(u)x^{+}(v)}=1.
\ea
A special solution valid for $L=S$ is 
\be
u=v=\left\{\begin{array}{ll}
\displaystyle \frac{1}{2}\cot\left(\frac{\pi}{2S}\right)+2\,\sin\left(\frac{\pi}{S}\right)\,g^{2}+\mathcal{O}(g^{4}) & \qquad {\rm any}\ S \\ \\
\displaystyle -\frac{1}{2}\tan\left(\frac{\pi}{2S}\right) -2\,\sin\left(\frac{\pi}{S}\right)\,g^{2}+\mathcal{O}(g^{4})& \qquad {\rm only\ for\ even}\ S.
\end{array}\right.
\ee
The wrapping correction can be computed and the first cases are as follows.

\medskip
\medskip
\noindent
\underline{\em\Large L = S = 2}
\medskip
\medskip

Let us choose $u_{4,1}^{(0)}=u_{4,2}^{(0)}=\frac{1}{2}\cot\frac{\pi}{2S} = \frac{1}{2}$. The other possible choice
$u_{4,1}^{(0)}=u_{4,2}^{(0)}=-\frac{1}{2}\tan\frac{\pi}{2S} = -\frac{1}{2}$
gives the same result. After some computation, we find
\be
Y_{a,0}^{*\ \rm LO}(u) = 
\frac{65536 a^2 \left((u-1) \left(a^2+4 u^2\right)+1\right)^2}{((a-2) a+4 (u-1) u+2)^2 (a (a+2)+4 (u-1) u+2)^2 \left(a^2+4 u^2\right)^4}
\,g^{8}.
\ee
Integrating over $u$ and summing over $a$, we obtain 
\ba
\delta E^{\rm LO} &=& 
\frac{64}{3}\,\left[9+12\,{\rm Im}\,\psi'\left(\frac{3}{2}+\frac{i}{2}\right)
+12\,{\rm Re}\,\psi''\left(\frac{3}{2}+\frac{i}{2}\right) \right. \\
&&\left.
-2\,{\rm Im}\,\psi^{(3)}\left(\frac{3}{2}+\frac{i}{2}\right)
+24\,\zeta_{3}-30\,\zeta_{5}
\right]\,g^{8},\nonumber
\ea
where $\psi(z) = (\log\Gamma(z))'$.

\medskip
\medskip
\noindent
\underline{\em\Large L = S = 3}
\medskip
\medskip

Taking $u_{4,1}^{(0)}=u_{4,2}^{(0)}=\frac{1}{2}\cot\frac{\pi}{2S}=\frac{\sqrt 3}{2}$, after some computation, we find the following wrapping correction
($\omega = e^{\frac{2\pi\,i}{3}}$)
\ba
\delta E^{\rm LO} &=& 
\frac{2}{3}\left[
15-63\,{\rm Re}\,\psi''(2+\omega)+28\,\sqrt{3}\,{\rm Im}\,\psi^{(3)}(2+\omega)
+7\,{\rm Re}\,\psi^{(4)}(2+\omega)\right. \\
&&\left. -138\,\zeta_{3}+840\,\zeta_{5}-840\,\zeta_{7}
\right]\,g^{10}.\nonumber
   \ea

\subsubsection{$N$-particle states for $L=2$ in the $\Z_{2}$ orbifold}

Another interesting class of states is obtained for $S=2$ and $L=2$ with an odd number of magnons $N$.
The twist constraint is automatically satisfied. The ciclicity condition reads
\be
\prod_{j=1}^{N} \frac{x^{+}_{4,j}}{x^{-}_{4,j}} = -1.
\ee
The Bethe equations are, as before, untwisted. It is clear that we find a solution by taking the same Baxter polynomial
as in Eq.~(\ref{eq:baxter2}). Indeed, for $t_{2}=0$ and even $N$, that polynomial solves the Bethe equations for 
cyclic states. For odd $N$, we obtain solutions of the Bethe equations with momentum $p=\pi$ which is 
precisely the modified ciclicity requirement. Working out the LO wrapping correction for the first cases of odd $N$,
we find ($N=1$ was already given in (\ref{eq:spin1}))
\ba
\delta E^{\rm LO}_{N=3} &=& \frac{484}{243}\,(215+744\,\zeta_{3}-1080\,\zeta_{5})\,g^{8}, \\ \nonumber \\
\delta E^{\rm LO}_{N=5} &=& \frac{18769}{4860000}\,(196147+579648\,\zeta_{3}-864000\,\zeta_{5})\,g^{8}.
\ea
In order to guess a closed formula as a function of $N$, one needs to compute the wrapping correction for large values
of $N$. The computation can be made very efficient by extending the formula (\ref{eq:wing}) to this case. To this aim,
we can use the following expansions 
\ba
\widetilde R^{(+)} &=& Q_{4}^{+}\,\left[
1+g^{2}\,\left(
-\frac{1}{2} (\log Q_{4})''_{u=i/2}-\frac{1}{u} (\log Q_{4})'_{u=i/2}\right)\right], \\
\widetilde R^{(-)} &=& Q_{4}^{-}\,\left[
1-g^{2}\,\left(
\frac{1}{2} (\log Q_{4})''_{u=i/2}-\frac{1}{u} (\log Q_{4})'_{u=i/2}\right)\right], \\
\widetilde B^{(+)} &=& Q_{4}(i/2)\,\left[
1+g^{2}\,\left(
\frac{1}{2} (\log Q_{4})''_{u=i/2}+\frac{1}{u} (\log Q_{4})'_{u=i/2}\right)\right], \\
\widetilde B^{(-)} &=& Q_{4}(-i/2)\,\left[
1+g^{2}\,\left(
\frac{1}{2} (\log Q_{4})''_{u=i/2}-\frac{1}{u} (\log Q_{4})'_{u=i/2}\right)\right].
\ea
Assuming that $Q_{4}$ is {\em odd}, one finds 
\ba
c &=& \sum_{j}\frac{1}{u_{4,j}+\frac{i}{2}}\frac{1}{u_{4,j}-\frac{i}{2}} = \left.
i\,(\log Q_{4})'\right|_{u=-i/2}^{u=+i/2}, \\
T_{a,0}^{*} &=& \left. i\,c\,g^{2}\,\frac{(-1)^{a+1}}{Q_{4}^{[1-a]}}
\mathop{\sum_{p=-a}^{a}}_{\Delta p = 2}\frac{Q_{4}^{[-1-p]}-Q_{4}^{[1-p]}}{u-p\frac{i}{2}} 
\right|_{Q_{4}^{[-a-1]} , Q_{4}^{[a+1]}\to 0}
\ea
This is formally the transfer matrix for an untwisted wing associated with an even $Q_{4}$ (see for instance
\cite{Beccaria:2010kd}
). 
In other words, for this orbifold, the non trivial phases $\lambda_{2}^{\ell, r} = \lambda_{4}^{\ell, r}=-1$
are compensated by the fact that $Q_{4}$ is odd. Technically, this comes from factors $Q_{4}(i/2)/Q_{4}(-i/2)=-1$
from the ratios of $\widetilde B$.

In conclusion, the wrapping correction is nothing but the Konishi correction evaluated in \cite{Bajnok:2008qj}
and applied here to the case of odd spin $N$. In other words, we simply have 
\ba
\delta E^{\rm LO}_{\mbox{\scriptsize odd}\ N} &=&
S_{1}^{2}\,\left[
256\,(
S_{-5}-S_{5}+2\,S_{-2,-3}-2\,S_{3,-2}+2\,S_{4,1}-4\,S_{-2,-2,1}
)\right. + \\
&& \left. -128\,(4\,S_{-2}\,\zeta_{3}+5\,\zeta_{5})\right]\,g^{8}, \nonumber
\ea
where all harmonic sums are evaluated at $N$. 

\section{Conclusions}
\label{sec:conclusions}

In this paper we have proposed a simple twisted Y-system for the computation of the leading order wrapping
corrections to general states in $\Z_{S}$ orbifolds of \sym. Our proposal is summarized in 
the twisted generating functions (\ref{eq:twistedgenerating}) , fusion factor (\ref{eq:twistedfusion}), and twist coefficients (\ref{eq:twistedlambda}). The computations are no more difficult that in the untwisted theory.
In particular, many states in the $\mathfrak{sl}(2)$ sector (with only central node excitations) can be treated
very explicitly. We recover all known results for physical states in a class of non-supersymmetric orbifolds treated 
in the TBA formalism of \cite{Arutyunov:2010gu,deLeeuw:2011rw}. Moreover, general orbifolds are included in the 
formalism and are under control as well.

Remarkably, in the non-supersymmetric case, the leading wrapping for $L=3$ twist operators is reciprocity 
respecting as the $L=2$ case precisely as it happened in the untwisted theory. It would be very interesting to pursue 
next-to-leading corrections to see whether reciprocity is broken or not. 

On the computational side, the proposed Y-system
cover the full set of states, not only those in the $\mathfrak{sl}(2)$ sector. Computing corrections in these cases
requires solving the one-loop Bethe equations as well as resumming the expansion (\ref{eq:twistedgenerating}).
Given the good convergence of the sums over intermediate states, this is something that is in principle accessible to 
numerics.

It would be also very important to confirm or disprove the proposed Y-system starting from the rigorous and general 
TBA formalism for the mirror string theory. This seems to be a mandatory step if one is interested in numerical investigations and strong coupling extrapolations or predictions. 
\section*{Acknowledgments}

We thank A. A. Tseytlin, R. Roiban, N. Gromov, F. Levkovich-Maslyuk, G. Arutyunov for helpful 
and stimulating discussions.

\appendix

\section{Gauge transformation of the $T$ functions}
\label{app:gauge}

The functional generator for $T$ functions takes the typical form 
\be
\sum_{a} D^{a}(T_{a} f_{a})\,D^{a} = \prod_{i} (1+D\,F_{i} D)^{n_{i}},
\ee
where $D=e^{-\frac{i}{2}\partial_{u}}$ and 
\be
f_{a} = \prod_{n=-\frac{a-1}{2}}^{\frac{a-1}{2}} f(u+i\,n).
\ee
Using $f(u+i\,n) = f^{[2n]}$, and $\ddb=e^{-i\,\partial_{u}}=D^{2}$ we can write 
\be
\sum_{a} D^{a}(T_{a} f_{a})\,D^{a} = D\prod_{i} (1+F_{i}\,\ddb)^{n_{i}}\,D^{-1}
\ee
\be
\sum_{a} D^{a-1}(T_{a} f_{a})\,D^{a+1} = \prod_{i} (1+F_{i}\,\ddb)^{n_{i}},
\ee
and shifting, 
\be
\sum_{a} T_{a}^{[-(a-1)]} f_{a}^{[-(a-1)]}\,\ddb^a = \prod_{i} (1+F_{i}\,\ddb)^{n_{i}}.
\ee
Now, let us multiply as in 
\be
\sum_{a} T_{a}^{[-(a-1)]} f_{a}^{[-(a-1)]}\,G^{-1}\,\ddb^a\,G = \prod_{i} (1+F_{i}\,G^{-1}G^{--}\ddb)^{n_{i}}.
\ee
Let us define
\be
U=G^{-1}G^{--}.
\ee
Then one can prove that 
\be
G^{-1}\,\ddb^a\,G = U_{a}^{[-(a-1)]}\,\ddb^{a}.
\ee
In other words, we have shown that 
\be
\sum_{a} T_{a}^{[-(a-1)]} (f\,U)_{a}^{[-(a-1)]}\,\ddb^a = \prod_{i} (1+F_{i}\,U\,
\ddb)^{n_{i}}.
\ee
This means that we can multiply by $U$ both $F_{i}$ and $f$. From these results, it is easy to 
recover the generating functional discussed in \cite{Gromov:2010dy}.


\end{document}